# تحلیل احساس در رسانه‌های اجتماعی فارسی با رویکرد شبکه عصبی پیچشی

مرتضی روحانیان[1]، کارشناس ارشد، مصطفی صالحی[2]، دانشیار، علی درزی[3]، استاد تمام، وحید رنجبر[4]، استادیار

1- دانشکده علوم و فنون نوین - دانشگاه تهران - تهران - ایران — rohanian@ut.ac.ir
2- دانشکده علوم و فنون نوین - دانشگاه تهران - تهران - ایران — mostafa_salehi@ut.ac.ir
3- دانشکده ادبیات و علوم انسانی - دانشگاه تهران - تهران - ایران — alidarzi@ut.ac.ir
4- دانشکده مهندسی کامپیوتر - دانشگاه یزد - یزد - ایران — vranjbar@yazd.ac.ir

**چکیده:** افزایش کاربری شهروندان از رسانه‌های اجتماعی (مانند توئیتر، فروشگاه‌های برخط و غیره) آن‌ها را به منبعی عظیم برای تحلیل و درک پدیده‌های گوناگون تبدیل کرده است. هدف تحلیل احساس استفاده از داده‌های بدست آمده از این رسانه‌ها و کشف گرایش‌های پیدا و پنهان کاربران نسبت به موجودیت‌های خاص حاضر در متن است. در کار حاضر ما با استفاده از شبکه عصبی پیچشی، که نوعی شبکه عصبی پیش‌خور است، به  تحلیل گرایش نظرات در رسانه‌های اجتماعی در دو و پنج سطح و با در نظر گرفتن شدت آن‌ها می‌پردازیم. در این شبکه عمل کانولوشن با استفاده از صافی‌هایی با اندازه‌های مختلف بر روی بردارهای جملات ورودی اعمال می‌شود و بردار ویژگی حاصل به‌عنوان ورودی لایه نرم بیشینه برای دسته‌بندی نهایی جملات بکار می‌رود. شبکه‌های عصبی پیچشی با پارامترهای مختلف با استفاده از معیار مساحت زیر منحنی و بر روی مجموعه داده جمع‌آوری شده از رسانه‌های اجتماعی فارسی ارزیابی شدند و نتایج بدست آمده نشان‌دهنده بهبود کارایی آن‌ها در گستره رسانه‌های اجتماعی نسبت به روش‌های سنتی یادگیری ماشین به‌خصوص بر روی داده‌ها با طول کوتاه‌تر هستند.

**واژه‌های کلیدی:** تحلیل احساس، رسانه‌های اجتماعی، شبکه عصبی پیچشی، شدت نظرات، متون کوتاه

# Convolutional Neural Networks for Sentiment Analysis in Persian Social Media

**Morteza Rohanian, MSc[1], Mostafa Salehi, PhD[2], Ali Darzi, PhD[3], Vahid Ranjbar, PhD**

1- Faculty of New Sciences and Technologies, University of Tehran, Tehran, Iran, Email: rohanian@ut.ac.ir
2- Faculty of New Sciences and Technologies, University of Tehran, Tehran, Iran, Email: mostafa_salehi@ut.ac.ir
3- Faculty of Literature and Humanities, University of Tehran, Tehran, Iran, Email: alidarzi@ut.ac.ir
1- Department of Computer Enigneering, Yazd University, Yazd, Iran, Email: vranjbar@yazd.ac.ir

**Abstract:** With the social media engagement on the rise, the resulting data can be used as a rich resource for analyzing and understanding different phenomena around us. A sentiment analysis system employs these data to find the attitude of social media users towards certain entities in a given document. In this paper we propose a sentiment analysis method for Persian text using Convolutional Neural Network (CNN), a feedforward Artificial Neural Network, that categorize sentences into two and five classes (considering their intensity)  by applying a layer of convolution over input data through different filters. We evaluated the method on three different datasets of Persian social media texts using Area under Curve metric. The final results show the advantage of using CNN over earlier attempts at developing traditional machine learning methods for Persian texts sentiment classification especially for short texts.

نام نویسنده مسئول: مصطفی صالحی
نشانی نویسنده مسئول: ایران، تهران، خیابان کارگر شمالی، بعد از پل جلال آل احمد، روبروی کوچه دهم، دانشکده علوم و فنون نوین، اتاق ۳۳۷.

# 1- مقدمه

دانستن و تحلیل نظر دیگران همواره جزء اساسی از فرآیند تصمیم‌گیری انسان‌ها در طول تاریخ بوده است. افراد در هر جامعه‌ای هنگام روبرو شدن با چالش‌های متفاوت از مشورت اعضای آن جامعه بهره گرفته‌اند. امروزه گسترش رسانه‌های اجتماعی به افراد جامعه با نگاه‌های مختلف فرصت داده تا در فضای عمومی نظرات خود درباره پدیده‌های گوناگون با هم به اشتراک بگذارند. 69 درصد کاربران بالغ اینترنت از رسانه‌های اجتماعی به‌عنوان محلی برای بحث درباره موضوعات مختلف و اطلاع از نظرات دیگران استفاده می‌کنند [1]. محتوای تولید شده از فعالیت در این رسانه‌ها، که 28 درصد از حضور برخط کاربران را شامل می‌شود [2]، می‌تواند منبع عظیم داده، برای تحلیل و درک رفتار افراد در مواجهه با پدیده‌های مختلف باشد.

تحلیل احساس شاخه‌ای از پردازش زبان طبیعی است که به تحلیل گرایش‌های مردم نسبت به موجودیت‌های خاص و ویژگی‌های مرتبط به آن‌ها، به‌صورت خودکار می‌پردازد. این موجودیت‌ها می‌توانند محصولات، سرویس‌ها، مجموعه‌ها، افراد، اتفاقات و یا موضوعات مختلف باشند. تحلیل احساس متمرکز بر نظراتی در زبان طبیعی است که به صورت مشخص یا ضمنی دارای جهت‌گیری منفی، خنثی یا مثبت هستند. جملاتی که دارای جهت‌گیری‌اند، جملات نسبی[1] نامیده می‌شوند که در مقابل جملات عینی[2] قرار دارند که واقعیات را بدون اعمال نظر گوینده بیان می‌کنند. در تحلیل احساس، ما به بررسی جملات نسبی و جملات عینی، که بیانگر اتفاقات و واقعیاتی با بار منفی یا مثبت هستند، می‌پردازیم.

در [3] نویسندگان این نظرات را در 5 مرتبه از نظر شدت جهت‌گیری قرار دادند: مثبت احساسی، مثبت منطقی، خنثی، منفی منطقی و منفی احساسی.

پژوهش‌های زیادی روش‌های معمول در یادگیری ماشین را در امر تحلیل احساس مورد بررسی قرار داده‌اند [4]. رویکردهای معمول در این پژوهش‌ها اغلب بر پایه الگوریتم‌های بانظارت[3] و ویژگی‌های استخراج شده به‌صورت غیر خودکار بوده است [5]-[7]. این گزینش ویژگی معمولا به صورت دستی انجام می‌شود و بسته به موضوع و نوع متن متفاوت است. به همین دلیل مدل‌ها وابسته به متن بوده و حالت کلی و عمومی[4] ندارند. روش‌های یادگیری عمیق در سال‌های اخیر به عنوان مجموعه روش‌هایی با تعمیم‌پذیری بالا مورد توجه پژوهشگران حوزه پردازش زبان طبیعی بوده‌اند و استفاده از آن‌ها در تحلیل احساس به‌خصوص برای زبان انگلیسی رایج شده است. امروزه الگوریتم‌های سنتی یادگیری ماشینی به مرور جای خود را به روش‌های یادگیری عمیق در تحلیل احساس می‌دهند. دلیل آن این است که این روش‌ها امکان این را دارند که بدون دخالت انسانی، ویژگی‌های پیچیده فراوانی درباره داده استخراج کنند. لازمه استفاده از مدل‌های یادگیری عمیق، داشتن داده آموزش کافی، زمان و منابع رایانشی مناسب برای آموزش درست مدل شبکه عصبی است [8]-[10].

ما در این مقاله برای تحلیل احساس متن فارسی، از شبکه‌های عصبی پیچشی[5] (CNN) که نوعی شبکه عصبی پیش‌خور[6] و چند لایه هستند، استفاده می‌کنیم. در این شبکه‌ها برای بدست آوردن خروجی به جای اتصال هر نورون لایه ورودی به لایه خروجی، بر روی داده ورودی (بردارهای کلمات حاصل از جاسازی کلمات[7]) با استفاده از صافی‌های مختلف عمل کانولوشن صورت می‌گیرد. از آن‌جا که شبکه‌های پیچشی توان استخراج ویژگی از واحدهای زبانی با طول‌های مختلف را دارند، استفاده از آن‌ها نتایج بهتری نسبت به روش‌های سنتی یادگیری ماشین برای زبان‌ها با منابع فراوان به‌بار می‌آورد [11]. در این پژوهش شبکه‌های عصبی پیچشی برای اولین بار برای تحلیل احساس زبان فارسی بر روی داده‌های جمع‌آوری شده از اخبار و توئیتر بکار رفته است. داده‌ها دارای دو نوع برچسب‌گذاری دو و پنج‌تایی هستند و دارای کاربری و طول متغیرند. نتایج بدست آمده در این مقاله نشان می‌دهد که این شبکه‌ها برای زبان‌ها با منابع محدود مثل فارسی، کارآیی بهتری نسبت به روش‌های سنتی یادگیری ماشین دارند. در دسته‌بندی داده‌های متنی با طول کم این بهبود نتایج تا 12 درصد رسیده است. مهم‌ترین دستاوردهای ما در این مقاله به‌صورت زیر است:

- استفاده از شبکه‌های عصبی پیچشی برای دسته‌بندی جملات در متون فارسی که با توجه به اطلاعات ما قبلا برای زبان فارسی انجام نشده است.

- تحلیل احساس در پنج سطح مختلف برای زبان فارسی و در نظر گرفتن شدت قطبیت.

- تحلیل احساس بر روی واحدهای زبانی با طول‌های متفاوت و بررسی روش پیشنهادی بر روی گستره‌ای از متون فارسی. در این راستا مجموعه داده تحلیل احساس با برچسب‌زنی جمع‌آوری شده از توئیتر (متون محاوره با طول‌های متفاوت) و سایت‌های خبری فارسی (متون رسمی) تهیه شده است.

در ادامه ما در بخش 2 به پژوهش‌های مرتبط با کار تحلیل احساس و در بخش 3 به معرفی روش پیشنهادی می‌پردازیم. در بخش 4 نتایج حاصل از روش پیشنهادی را گزارش می‌کنیم و درباره آن‌ها بحث می‌کنیم و بخش 5 به نتیجه‌گیری و کارهای آتی اختصاص دارد.

## 2- کارهای مرتبط

بیشتر مطالعات مرتبط با تحلیل احساس در گذشته بر اساس الگوریتم‌های یادگیری بانظارت انجام گرفته است که نیاز به تهیه داده برچسب خورده دارند. مدل بیز ساده[8]، ساده‌ترین و پراستفاده‌ترین الگوریتم احتمالاتی برای دسته‌بندی است و بر مبنای قضیه بیز کار می‌کند. این مدل احتمالات پسین رویدادها را محاسبه کرده و برچسبی که بیشترین احتمال پسین را دارد به رویداد نسبت می‌دهد. دسته‌بندی‌کننده

پرکاربرد دیگر آنتروپی بیشینه[9] است. آنتروپی بیشینه مدل احتمالاتی است که کار دسته‌بندی را می‌توان با آن انجام داد. این روش بر پایه مدل نمایی[10] و اصل حداکثرآنتروپی[11] است [12]. استفاده از این روش تجربه‌های موفقی در کار پردازش زبان طبیعی از جمله در تحلیل احساس به ارمغان آورده است [13]. این روش در اکثر (و نه در همه) مواقع نسبت به مدل بیز ساده برتری دارد [4]. ماشین بردار پشتیبان[12] (SVM) برای کار دسته‌بندی اسناد بر مبنای موضوعات مشابه بسیار مفید است [14]. روش SVM یک مدل یادگیری بانظارت است که کار آن دسته‌بندی کردن اشیا در کلاس‌های مختلف با استفاده از ویژگی‌های استخراج شده است. این دسته‌بندی، با ایجاد ابرصفحه‌ای میان نمونه‌های هر کلاس و حداکثر کردن فاصله نمونه‌ها از این صفحه صورت می‌گیرد [32]. برتری این روش نسبت به دیگر روش‌های مطرح یادگیری ماشین آن است که در مورد داده‌های ورودی پیش‌فرضی ندارد و به جای تکیه بر ارزش‌های احتمالاتی، سعی دارد تا بهینه‌ترین دسته‌بندی را با داده‌های موجود انجام دهد و نتایج بدست آمده از آن در تحلیل احساس برتری محسوسی به دیگر روش‌های یادگیری ماشین در زبان انگلیسی دارد [6].

در سال‌های اخیر روش‌های یادگیری عمیق به‌خصوص شبکه‌های عصبی بازگشتی[13] (RNN) در تحلیل احساس برای زبان انگلیسی [15]، چینی [16] و آلمانی [17] در میان زبان‌های مختلف، با استفاده از بردارهای مختلف نمایش کلمات کاربرد زیادی داشته است. آن‌ها برای درک و کنترل ترکیب معنایی در کارهای پیچیده‌ای مانند تحلیل احساس مفید هستند. شبکه‌های RNN برای داده‌هایی با قابلیت تبدیل به مقادیر متوالی به کار می‌روند و با استفاده از ایده‌ی اشتراک گذاری پارامترها برای رسیدن به وزن‌های مطلوب، توانایی پردازش توالی‌هایی با طول‌های متفاوت را دارند [18]. با وجود این‌که استفاده از آن‌ها در تحلیل احساس برای زبان انگلیسی با نتایجی بهتر از روش‌های یادگیری بانظارت همراه بوده است [9]، با رشد ساختار شبکه‌های RNN، ابعاد ماتریس‌ها در مرحله بازپخش به صورت توانی رشد می‌کنند و در عمل استفاده از آن‌ها غیرممکن می‌شود [33].

شبکه‌های پیچشی که کولوبرت و دیگران [19] در ابتدا برای کاربرد در بینایی رایانه‌ای ارائه کرده‌اند، اخیرا در بسیاری از کارهای پردازش زبان طبیعی مانند تجزیه نحوی، تجزیه سطحی، برچسب زنی نقش معنایی[14]، و قطعه‌بندی[15] مورد استفاده قرار گرفته است. استفاده از شبکه‌های پیچشی در تحلیل احساس نیز برای زبان‌هایی با منابع فراوان مورد استفاده قرار گرفته و باعث بهبود قابل توجه دقت و کاهش زمان مرحله آموزش نسبت به دیگر روش‌های یادگیری عمیق شده است [10].

پژوهش‌های حوزه تحلیل احساس در زبان فارسی معمولا یا با استفاده از روش‌های مبتنی بر قاعده هستند یا مبتنی بر پیکره [24]. برای بهبود نتایج معمولا از پیش پردازش نظرات و ویژگی‌های لغتنامه استفاده شده است [30]. بصیری و همکاران [28] یک چارچوب مبتنی بر لغتنامه ارائه کردند که به صورت بدون نظارت با استفاده از قواعد از پیش تعیین شده و لغتنامه تعریف شده جهت‌گیری متون محاوره را تشخیص می‌دهد. استفاده از SVM برای تحلیل احساس در زبان فارسی بر روی داده مربوط به نقد فیلم، منجر به نتایج بهتری نسبت به روش‌های دیگر یادگیری ماشین شده است [27]. بازدهی این روش‌ها وابسته به کیفیت برچسب‌دهی در پیکره‌ها و شیوه گزینش ویژگی‌ها پیش از شروع کار دسته‌بندی است. روشنفکر و همکاران [29] برای اولین بار از شبکه‌های عصبی LSTM برای تشخیص احساس متون فارسی استفاده کردند و توانستند نسبت به روش‌های یادگیری سنتی نتایج بهتری داشته باشند، اما این نوع شبکه‌ها برای آموزش نیاز به داده‌های خیلی زیادی هستند. همچنین آن‌ها در کار خود فقط دو سطح از احصاص را در نظر گرفتند و از جاسازی ساده کلمات استفاده کردند.

به‌طور کلی مزایای استفاده از یادگیری عمیق شامل موارد زیر است [20]:

- احتیاجی نیست ویژگی‌ها به صورت دستی تهیه شوند. در یادگیری عمیق به جای استخراج دستی ویژگی‌ها معمولا از جاسازی کلمات استفاده می‌شود که در آن‌ها اطلاعات مربوط به بافت متنی وجود دارند.
- با استفاده از شبکه‌های عصبی یادگیری، انتخاب ویژگی‌ها[16] و نمایش آن‌ها می‌تواند هم با یادگیری بانظارت و هم بدون نظارت[17] صورت گیرد.
- در تحلیل احساس با متن‌های گوناگونی از لحاظ سبک نوشتار و بافت معنایی روبرو هستیم. انعطاف و تعمیم‌پذیری روش‌های یادگیری عمیق، اجازه می‌دهد تا با مشکل عدم تعمیم پذیری مدل کم‌تر روبرو شویم.

## ۳- راهکار پیشنهادی

قبل از استفاده از داده‌ها در روش پیشنهادی، ابتدا توسط ابزارهای موجود، پیش‌پردازش‌هایی نظیر بهنجارسازی، توکن‌بندی و جداسازی بر روی داده‌های ورودی انجام می‌شود. همچنین برای تحلیل احساس زبان‌ها با منابع محدود مثل فارسی با استفاده از یادگیری عمیق نیاز به مجموعه‌ای از بردارها برای نمایش کلمات داریم که آن‌ها را با استفاده از جاسازی کلمات روی مجموعه ویکی‌پدیای فارسی بدست می‌آوریم. این بردارها به‌عنوان داده ورودی شبکه برای استخراج ویژگی‌ها به‌کار می‌روند.

شبکه‌های عصبی پیچشی (CNN) نوعی خاص از شبکه‌های عصبی برای پردازش داده هستند که با بردارهای کلمات مانند یک تور[18] برخورد می‌کنند [10]. صافی‌های[19] هر لایه کانولوشن بر روی طول ماتریس‌های حاصل از بردارهای ورودی حرکت می‌کند. عرض صافی‌ها به اندازه عرض بردار ورودی (بعد بردار کلمات) و طول آن‌ها معمولا بین ۲ تا ۵ کلمه است. از بردارهای حاصل نگاشت‌های ویژگی[20] حاصل می‌شوند که با استفاده از لایه الحاق حداکثری[21] تبدیل به یک بردار نهایی می‌شوند و از آن به‌عنوان ورودی لایه آخر برای دسته‌بندی جملات استفاده می‌شود.

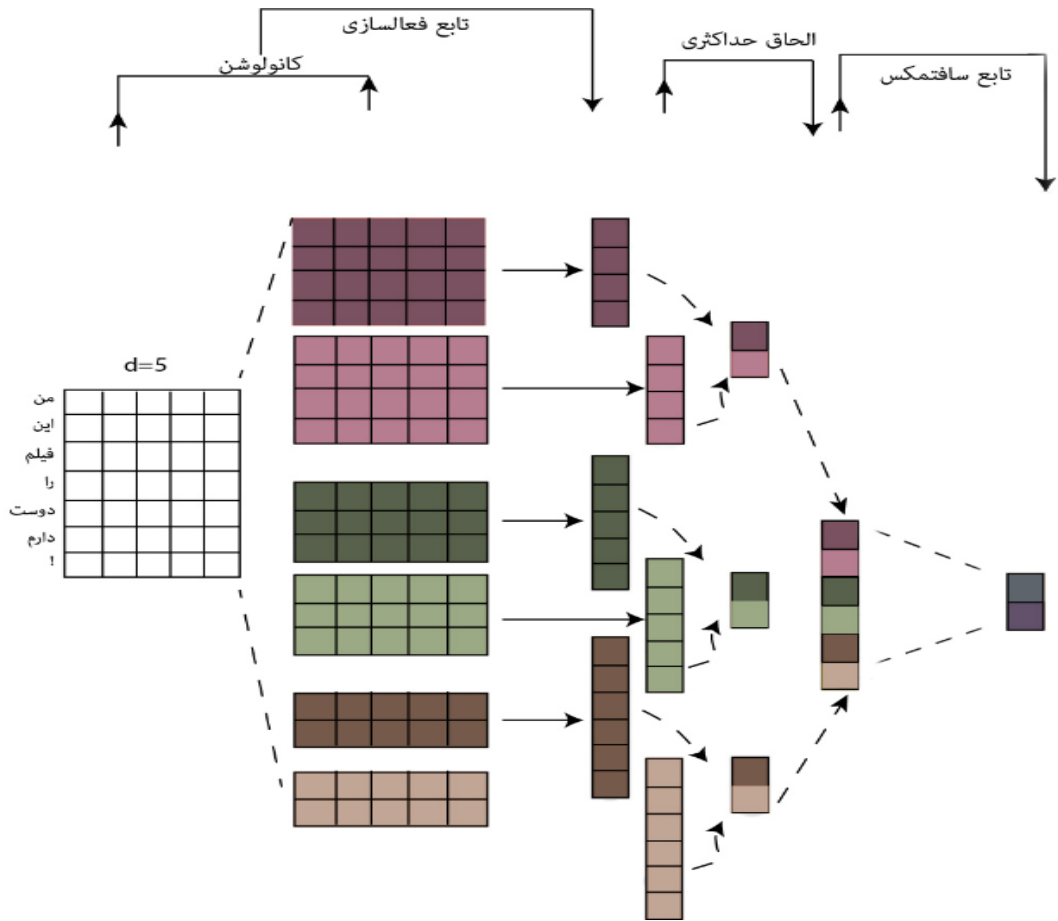

شکل ۱ – نمایش یک شبکه عصبی پیچشی برای دسته‌بندی جملات.

## ۱-۳- کانولوشن

پایه اصلی شبکه پیچشی بردارهای کلمات ورودی $x \in R^s$ هستند که در آن $s$ ابعاد بردارها است و هر سند ورودی به‌صورت ماتریس $d \in R^{n \times s}$ نمایش داده می‌شود که $n$ تعداد کلمات آن است و هر سطر ماتریس، بردار یک کلمه را نمایندگی می‌کند.

لایه کانولوشن که هدف از آن استخراج ویژگی‌های محلی از رشته‌های حرفی داخل جمله است، تعمیمی از رویکرد پنجره است که در آن چند صافی با اندازه مشخص کل جمله را می‌پیمایند و نتایج حاصل از کانولوشن روی ماتریس‌های مختلف را با هم ترکیب می‌کنند. عرض این صافی‌ها به اندازه ماتریس ورودی و طول آنها قابل تعیین است. هر کدام از این صافی‌ها همان‌طور که در شکل ۱ مشخص است، یک ماتریس هستند که مجموع حاصل‌ضرب عناصر آن و ماتریس ورودی، ماتریس جدیدی است که بردار کلمه ورودی را با توجه به بافتش تغییر می‌دهد. ما یک لایه کانولوشن $W \in R^{h \times s}$ تعریف می‌کنیم که در آن $h$ تعداد کلماتی است که می‌خواهیم روی آن کانولوشن انجام دهیم (طول صافی). اگر کانولوشن را با عملگر $*$ نشان دهیم هر بار پیمایش صافی روی بردار برابر است با:

$$W * d_{j:j+h-1} = \sum_{i=j}^{j+h-1} \sum_{k=0}^{s-1} W_{i,k} d_{i,k}. \quad (1)$$

سپس صافی به طول $h$ کلمه، $d_{j:j+h-1}$، را با استفاده از تابع غیرخطی $f$ به عدد حقیقی $c_j$ نگاشت می‌کنیم:

$$c_j = f(d_{j:j+h-1} + b), \quad (2)$$

که $b$ در آن یک عدد حقیقی است که میزان تمایل[22] را نشان می‌دهد. با اعمال کانولوشن بر روی تمام سند با استفاده از $W$، یک بردار ویژگی حاصل می‌شود

$$c(W) = [c_1, c_2, \ldots, c_{n-h+1}]. \quad (3)$$

هر شبکه می‌تواند دارای مقادیر متفاوت برای نوع صافی‌ها و ماتریس‌های وزنی باشد که به هر کدام کانال[23] گفته می‌شود. در حین مرحله آموزش یک شبکه عصبی پیچشی بر پایه کاربری خاص ایجاد می‌شود که عناصر ماتریس‌های صافی یاد گرفته می‌شوند. چون هر عنصر ورودی و صافی باید به طور جداگانه ذخیره‌سازی شوند، معمولا فرض می‌کنیم که این عناصر جز در نقاط محدودی که مقادیر آنها ذخیره شده‌اند، در بقیه نقاط دارای مقدار صفر هستند.

## ۲-۳- الحاق حداکثری

هدف از لایه الحاق حداکثری، یکسان‌سازی طول بردارهای جملات[24] و کاهش ابعاد بردار خروجی در عین حفظ اطلاعات مهم است. برای مثال اگر ۵۰۰ صافی وجود داشته باشد، بعد از اعمال الحاق حداکثری، برداری

۵۰۰ بعدی که داریم که به‌عنوان ورودی با طول ثابت، برای بیشینه‌نرم[۲۵] استفاده می‌شود. همچنین اگر هر صافی را دارای اطلاعات مربوط به یک ویژگی خاص ورودی بدانیم با استفاده از الحاق حداکثری می‌توانیم، پیش‌بینی کنیم که آیا این ویژگی در جمله وجود داشته یا نه. کار الحاق حداکثری در این لایه این است که بیشترین مقدار هر ویژگی را از میان صافی‌های مختلف برگزیند

$$\hat{c}_W = \max c(W)_i. \qquad (4)$$

این روش نسبت به میانگین‌گیری بهتر است چرا که در دسته‌بندی، همه کلمات به یک اندازه مهم نیستند و اهمیت نسبی آن‌ها در لایه الحاق حداکثری لحاظ می‌شود. در نهایت از همه صافی‌ها یک بردار ویژگی سراسری[۲۶] بدست می‌آید که ورودی لایه بعد محسوب می‌شود:

$$\hat{C}_W = [\hat{c}_{W^1}, ..., \hat{c}_{W^k}]^T, \qquad (5)$$

در رابطه ۵، $T = \{1, ..., k\}$ است. اندازه بردار ویژگی سراسری برای جملات مختلف ثابت است. در لایه الحاق، ما عمل الحاق حداکثری را برای ترکیب ویژگی‌های مختلف حاصل از لایه کانولوشن و به وجود آوردن یک بردار با بعد ثابت انجام می‌دهیم. برای رسیدن به بردار جمله خروجی صافی‌های لایه کانولوشن را پیوند زنجیره‌ای می‌دهیم. هرچه تعداد صافی‌ها بیشتر باشد، تعداد نگاشت‌های ویژگی بیشتر می‌شود و از هر نگاشت ویژگی، بیشترین مقدار در مرحله الحاق حداکثری انتخاب می‌شود و به بردار ویژگی سراسری اضافه می‌شود. پس می‌توان گفت که اندازه این بردارها به تعداد نگاشت‌های ویژگی و تعداد کانال‌ها وابسته است. در این مرحله اطلاعات مربوط به مکان قرارگیری ویژگی‌های مختلف را با توجه به در نظر نگرفتن ترتیب کلمات از دست می‌دهیم.

### ۳-۳- بیشینه‌نرم

در لایه آخر از تابع بیشینه‌نرم، که نوع گسترش داده شده رگرسیون لجستیک است، برای دسته‌بندی بردار ویژگی بدست آمده استفاده می‌کنیم. اگر $U \in R^{2 \times k}$ و $b^U \in R^2$ پارامترهای لایه نرم بیشینه باشند و ورودی وزن‌دار برابر باشد با:

$$y_j = U_j \hat{c}_W + b_j^U \qquad (6)$$

که در آن $c_w$ بردار ورودی، $U_j$ ردیف $j$-ام $U$ و $b_j^U$ عنصر $j$-ام $b^U$ و $Y_j$ برچسب $j$-ام در ماتریس $d$ است. اندازه این لایه برابر با تعداد برچسب‌ها است. احتمال برچسب خروجی برابر است با:

$$P(Y_j = 1|d, W, b^U) = \frac{e^{y_j}}{\sum_i e^{y_i}}, \qquad (7)$$

اگر $D$ مجموعه ماتریس‌های داده آموزش باشد، برای آموزش دسته‌بندی دوتایی باید روابط (۸) و (۹) را به ترتیب برای داده‌های منفی و مثبت کمینه کرد:

$$-\sum_{d \in D} \log(P(Y_{POS}^d - 1|d, W, b^U)) \qquad (8)$$

$$-\sum_{d \in D} \log(P(Y_{POS}^d - 2|d, W, b^U)) \qquad (9)$$

پارامترهای شبکه عصبی پیچشی $(d, W, b^U)$ که روابط (۸) و (۹) را کمینه می‌کنند با محاسبه گرادیان از طریق روش پس‌انتشار بدست

می‌آیند. شکل کلی معماری CNN پیشنهادی در شکل ۱ قابل مشاهده است.

### ۳-۴- جاسازی کلمات

بر خلاف بسیاری از روش‌های دیگر در یادگیری ماشین، ورودی در اینجا به صورت متنی نیست. این به آن معناست که برای تهیه ورودی، متن باید به بردارهای ویژگی یا به عبارت دیگر به جاسازی‌های کلمات تبدیل شوند که این جاسازی‌ها، اگر درست استخراج بشوند خود حاوی اطلاعات بافتی و معنایی متن هستند [۲۱].

این بردارها با آموزش دادن شبکه عصبی بر مبنای پیکره متنی حاصل می‌شوند و فرایندی زمان‌بر هستند. با استفاده از یادگیری عمیق، یک مدل زبانی برای یادگیری نمایش توزیع یافته کلمات[۲۷] بر اساس سه ایده کلی زیر می‌توان ارائه کرد [۲۲]:

- هر کلمه در پیکره به یک بردار ویژگی $S$ بعدی حاوی اعداد حقیقی متناظر می‌شود.
- تابع احتمال توام برای کلمات، با استفاده از این نمایش‌های برداری بیان می‌شود.
- یادگیری بردارهای ویژگی کلمات و پارامترهای تابع احتمال به طور همزمان انجام می‌شوند.

هر کدام از جاسازی‌های کلمات می‌تواند ابعادی به دلخواه کاربر داشته باشد. بعد بالاتر به این معنی است که اطلاعات بیشتری ضبط شده است ولی در عین حال با زیاد شدن بعد، هزینه‌های محاسباتی نیز افزایش می‌یابد.

## ۴- ارزیابی راهکار پیشنهادی

برای ارزیابی راهکار پیشنهادی و مقایسه‌ی نتایج آن با روش‌های سنتی یادگیری ماشین، آن را با پارامترهای مختلف بر روی مجموعه داده‌ی رسانه‌های اجتماعی فارسی آزمایش می‌کنیم.

### ۴-۱- مجموعه داده

برای آموزش و ارزیابی مدل از مجموعه داده‌های مختلف استفاده شده است:

جدول ۱: الگوهای مشخص برای توییت‌ها

| نتیجه | مثال | محتوا |
|---|---|---|
| <کاربر> | @username | نام کاربری |
| <آدرس> | http://t.co/url | URLs |
| <عدد> | 2007/۱۳۸۹ | اعداد |
| <هشتگ> | #tweet | هشتگ‌ها |
| یا | ∧ | خط مورب |

جدول ۲: تعداد توکن‌ها و پراکندگی آن‌ها در توییت‌های مجموعه داده توئیتر فارسی پس از برچسب زنی دو سطحی و پنج سطحی

| ویژگی‌ها | پنج‌تایی | | | | | دوتایی | |
|---|---|---|---|---|---|---|---|
| | منفی احساسی | منفی منطقی | خنثی | مثبت منطقی | مثبت احساسی | منفی | مثبت |
| تعداد توییت | ۱۶۴۷ | ۱۵۸۶ | ۳۲۵۴ | ۱۶۹۴ | ۱۷۸۲ | ۵۸۶۳ | ۵۲۹۷ |
| تعداد توکن | ۲۶۸۴۶ | ۲۵۲۶۹ | ۵۱۰۹۸ | ۲۷۸۹۵ | ۲۷۶۵۸ | ۹۳۸۰۵ | ۸۴۹۴۵ |
| میانگین توکن بر توییت | ۱۶.۲۹ | ۱۵.۹۳ | ۱۵.۷ | ۱۶.۴۶ | ۱۵.۵۲ | ۱۵.۹۹ | ۱۶.۰۳ |
| تعداد واژگان | ۵۲۷۰ | ۵۱۰۵ | ۷۱۲۱ | ۵۶۷۸ | ۵۹۱۲ | ۷۸۶۵ | ۷۰۸۹ |

- مجموعه داده سنتی‌پرس: این مجموعه ۱۱۰۰ نظر درباره محصولات است که از فروشگاه برخط دیجی‌کالا جمع‌آوری شده، که شامل جملات فارسی با برچسب‌های حاوی بار معنایی است که در پردازش زبان طبیعی و به طور مشخص در زمینه تحلیل احساس یا عقیده‌کاوی کاربرد دارد [۳۱].

- مجموعه داده توئیتر فارسی: ۱۱۱۶۰ توئیت جمع‌آوری شده و برچسب خورده در دو و پنج سطح برای ارزیابی و آموزش در بازه بین فروردین تا تیر ۱۳۹۶ (برچسب‌ها به صورت دستی و توسط افراد متخصص زبان فارسی ایجاد شده است).

- مجموعه داده اخبار: ۱۶۴۳ خبر فارسی از ۳۰ منبع خبر فارسی در بازه بین فروردین تا تیر ۱۳۹۶ جمع‌آوری شده و خبرها نیز به صورت دستی برچسب‌خورده است. برای مجموعه داده توئیتر، ۱۵۰۰۰ توییت از سایت شبکه اجتماعی توئیتر جمع‌آوری شد. ابتدا برای پردازش به آرایه‌ای از توکن‌ها تبدیل شدند. بدلیل اینکه ما تنها به تحلیل زبان فارسی می‌پردازیم حروف زبان‌های دیگر از متون حذف شدند. توییت‌ها با الگوی مشخص شده در جدول ۱ اصلاح شدند. استفاده از این الگوها برای یکسان‌سازی توکن‌های داده‌ها و بردارهای پیش آموخته بوده است.

توییت‌ها و اخبار با استفاده از ۳ نفر و در پنج دسته برچسب‌گذاری شدند. داده‌هایی که حداقل ۲ نفر به آن یک برچسب داده بودند برای آموزش و ارزیابی پنج‌تایی در نظر گرفته شدند. اگر ۲ نفر به یکی از داده‌ها برچسبی از یک نوع جنسیت ولی با شدت متفاوت داده بودند (مثلا مثبت احساسی و مثبت منطقی)، آن داده برای دسته‌بندی دوتایی بکار گرفته شد. در نهایت ۹۹۶۳ توییت و ۱۱۵۴ خبر برای دسته بندی پنج‌تایی و ۱۱۱۶۰ توییت و ۱۶۴۳ خبر برای دسته‌بندی دوتایی باقی ماند. در جدول ۲ اطلاعات مربوط به تعداد توکن‌ها و پراکندگی آن‌ها در دسته‌های مختلف نمایش داده شده است.

ما از بردارهای پیش‌آموخته ویکی‌پدیای فارسی برای دستیابی به بردارهای جاسازی کلمات استفاده کردیم که بر روی تعداد هم‌آیی کلمات در متن، آموزش دیده شده است [۲۳].

۲۰ درصد هر کدام از مجموعه داده‌ها به‌عنوان داده ارزیابی و ۸۰ درصد بقیه به عنوان داده آموزش در نظر گرفته شده است.

### ۲-۴- پیاده‌سازی

برای استفاده کامل از منابع محاسباتی GPU، پیاده‌سازی در محیط Keras و Theano [۲۶] که چارچوبی سطح بالا در زبان پایتون برای پیاده‌سازی شبکه‌های عصبی عمیق است، صورت گرفت. برای تنظیم پارامترها از تابع جستجوی شبکه‌ای scikit-learn [۲۵] که می‌تواند با استفاده از تمام ترکیبات پارامترهای احتمالی بهترین عملکرد را شناسایی کند، استفاده شده است. آموزش همراه با توقف اولیه صورت گرفته است. به این معنا که اگر زیان اعتبارسنجی[۲۸] بعد از پنج دور افزایش نیاب، پردازش متوقف می‌شود. برای آموزش همه مجموعه داده‌ها از صافی‌هایی سه‌تایی و ابعاد ۱۰۰ برای بردارهای ورودی استفاده شده است. در مرحله آموزش تعداد تکرار برابر ۱۰۰ در نظر گرفته شد. تعداد دسته‌ها دو و پنج در نظر گرفته شده و برای تنظیم کردن از دراپاوت[۲۹] با نرخ ۰.۵ در لایه ماقبل آخر استفاده شده است. آموزش با استفاده از گرادیان کاهشی اتفاقی و قانون بهروز رسانی آدادلتا[۳۰] صورت گرفته است. برای رسیدن به تحلیل مناسب شرایط تصادفی (مقداردهی اولیه برای شبکه و جاسازی کلمات و دیگر پارامترها) برای همه مدل‌ها یکسان در نظر گرفته شده است.

### ۳-۴- معیار ارزیابی

برای ارزیابی از مساحت سطح زیر نمودار دو بعدی که در آن نرخ تشخیص صحیح دسته مثبت روی محور Y و نرخ تشخیص غلط دسته منفی روی محور X رسم می‌شود استفاده می‌کنیم که به آن مساحت زیر منحنی[۳۱] (AUC) می‌گوییم. هرچه عدد زیر نمودار بزرگتر باشد دسته‌بندی صورت گرفته دقیق‌تر بوده است.

### ۴-۴- نتایج

نتایج مدل‌های ما با پارامترهای متفاوت برای دو و پنج دسته در مقایسه با مدل‌های یادگیری ماشین سنتی و شبکه‌های عصبی بازگشتی در جدول ۳ آمده است. ارزیابی‌ها با معیار مساحت زیر نمودار نشان

می‌دهند که مدل شبکه‌های پیچشی به‌طور قابل ملاحظه‌ای عملکرد بهتری در هر دو دسته نسبت به روش‌های پراستفاده سنتی یادگیری ماشین و شبکه‌های عصبی بازگشتی دارند. تعدد لایه‌ها برای پردازش، که ویژگی‌های سطح بالا را برخلاف روش‌های سنتی یادگیری ماشین بدون نظارت استخراج می‌کنند و وجود لایه الحاق حداکثری که نمونه‌های مناسبی از ویژگی‌ها را انتخاب می‌کند، به بهبود نتایج کمک کرده‌اند. با وجود این‌که برخلاف بسیاری روش‌های دیگر یادگیری ماشین شبکه عصبی پیچشی ترتیب کلمات را محسوب نمی‌کند، صافی‌های موجود شبکه تا سه کلمه و بیشتر را در کنار هم در نظر می‌گیرند که محاسبه آن در روش‌های سنتی بسیار دور از ذهن است. این موضوع به‌خصوص در جملاتی که در آن‌ها، بین بخشی از عبارتی خاص با قرار گرفتن کلمات دیگر فاصله افتاده بکار می‌آید.

به‌طور کلی افزایش بعد و تعداد تکرار در مرحله آموزش بردارهای ورودی کلمات به بهتر شدن کارایی مدل منجر می‌شود. این بردارها که از متون ویکی‌پدیای فارسی بدست آمده‌اند، توانایی ذخیره کردن اطلاعات مربوط به روابط معنایی موجود در بردارهای کلمات را دارند، که محصول نمایش توزیع یافته آن‌هاست. این ویژگی توانایی شبکه عصبی پیچشی را برای کشف روابط معنایی ترکیبی اجزای جمله نسبت به روش‌های دیگر یادگیری ماشین افزایش می‌دهد.

**جدول 3- نتایج ارزیابی مدل‌ها برای دسته‌بندی 2 و 5تایی روی داده‌های مختلف با معیار مساحت زیر نمودار**

| مدل | پنج‌تایی | | | دوتایی | | |
|---|---|---|---|---|---|---|
| | توئیتر | اخبار | نظرات | توئیتر | اخبار | نظرات |
| CNN | ۰.۴۳ | ۰.۴۶ | ۰.۴۳ | ۰.۷۵ | ۰.۷۹ | ۰.۷۳ |
| RNN | ۰.۳۹ | ۰.۴۱ | ۰.۴۰ | ۰.۶۳ | ۰.۷۴ | ۰.۶۹ |
| بیز ساده | ۰.۳۹ | ۰.۴۰ | ۰.۳۸ | ۰.۵۲ | ۰.۶۲ | ۰.۵۶ |
| آنتروپی بیشینه | ۰.۳۹ | ۰.۳۹ | ۰.۳۸ | ۰.۶۱ | ۰.۷۱ | ۰.۶۹ |
| SVM | ۰.۴۰ | ۰.۴۰ | ۰.۴۱ | ۰.۶۳ | ۰.۷۴ | ۰.۶۸ |

هم‌چنین نتایج بدست آمده نشان‌دهنده کارایی بالاتر شبکه‌های عصبی پیچشی بر روی رسانه‌های اجتماعی با طول متن کوتاه‌تر (توئیتر) نسبت به روش‌های سنتی یادگیری ماشین و شبکه‌های عصبی بازگشتی است. اگرچه عملکرد مدل پیشنهادی در محیط زبان غیررسمی و با علائم نگارشی و غیر نگارشی مرسوم با افت روبرو می‌شود اما در تحلیل متون خبری، بردارهای عمومی بدست آمده از ویکی‌پدیای فارسی، به‌دلیل استفاده از زبان رسمی، کارایی بهتری نسبت به دو داده دیگر نشان داده‌اند.

بدیهی است که علاوه بر بردارهای کلمات عمومی، استفاده از بردارهای کارویژه (در اینجا بردارهای مخصوص تحلیل احساس) به افزایش دقت مدل کمک خواهد کرد. همین‌طور در کار حاضر برای بدست‌آوردن بردارهای جملات ما از تجمیع بردارهای کلمات استفاده کردیم. برای بهبود کیفیت نمایش برداری متن می‌توان از روش‌های دیگری چون بردارهای پاراگراف، که قادر است متون با طول‌های مختلف را نمایندگی کند، بهره گرفت.

### 5-4- تاثیر تعداد صافی‌ها بر نتایج

تاثیر اندازه صافی را بر روی داده‌های مختلف و با تنظیم بعد بردارها بر روی ۱۰۰ و با ۱۰۰ بار تکرار اندازه گرفته شده‌است. اندازه‌های صافی را برابر دو، سه، پنج، هفت و نه می‌گیریم. نتایج برای دو و پنج دسته در جدول ۴ آمده است.

**جدول 4- نتایج ارزیابی مدل‌ها برای دسته‌بندی 2 و 5تایی روی داده‌های مختلف با در نظر گرفتن تاثیر تعداد صافی‌ها**

| اندازه صافی | پنج‌تایی | | | دوتایی | | |
|---|---|---|---|---|---|---|
| | توئیتر | اخبار | نظرات | توئیتر | اخبار | نظرات |
| دو | ۰.۳۹ | ۰.۴۳ | ۰.۴۴ | ۰.۷۴ | ۰.۶۸ | ۰.۶۸ |
| سه | ۰.۴۰ | ۰.۴۳ | ۰.۴۶ | ۰.۷۵ | ۰.۷۹ | ۰.۶۹ |
| پنج | ۰.۴۱ | ۰.۴۲ | ۰.۴۵ | ۰.۶۹ | ۰.۷۹ | ۰.۷۲ |
| هفت | ۰.۴۰ | ۰.۴۲ | ۰.۴۶ | ۰.۶۳ | ۰.۷۹ | ۰.۷۳ |
| نه | ۰.۴۰ | ۰.۴۰ | ۰.۴۳ | ۰.۶۳ | ۰.۷۹ | ۰.۷۲ |

نتایج بدست آمده نشان می‌دهد که مدل با صافی‌های دو و سه بر روی داده توئیتر بهتر عمل می‌کند. با افزودن بر اندازه صافی‌ها کیفیت مدل بر روی توئیتر بهبود پیدا نمی‌کند. با افزایش طول جملات به‌خصوص در دسته‌بندی پنج‌تایی، نیاز به صافی‌های بزرگ‌تر احساس می‌شود. عملکرد مدل برای دسته‌بندی نظرات و اخبار با بزرگ شدن صافی و رسیدن به اندازه پنج در هر دو بخش دو و پنج‌تایی رشد می‌کند. با تغییر صافی به هفت و نه بر روی داده نظرات و اخبار بهبود محسوسی پیدا نمی‌کند. مشاهدات ما نشان می‌دهد که بهتر است پیش از آموزش، مدل صافی با اندازه مناسب از راه آزمون و خطا بدست آید. در هر مدل هر صافی را می‌توان به تنهایی یا به‌صورت ترکیبی با صافی‌های دارای اندازه نزدیک بکار برد.

## 4-6- تاثیر ابعاد بردارهای کلمات بر نتایج

با قرار دادن اندازه صافی روی سه و تعداد تکرار ۱۰۰، ما ابعاد را برابر اعداد ۱۰، ۵۰، ۱۰۰، ۲۰۰ و ۳۰۰ قرار دادیم. بهترین عدد برای بعد انتخابی به نوع مجموعه داده بستگی دارد.

از نتایج نشان‌داده شده در جدول ۵ مشاهده می‌شود که عملکرد مدل با افزایش ابعاد بردارها به بیش از ۱۰ بهبود چشم‌گیری پیدا می‌کند. این موضوع بر رود داده اخبار که بیشترین طول را دارد محسوس‌تر است. می‌توان نتیجه گرفت افزایش ابعاد به بیش از ۲۰۰ در کار حاضر تاثیر چندانی در نتایج ندارد و حتی ممکن است کارآیی را کاهش دهد. همین‌طور با افزایش ابعاد بردارهای کلمات زمان مورد نیاز برای آموزش آن‌ها به طور قابل توجهی افزایش می‌یابد. در کاربرد به نظر می‌رسد تعداد دسته‌ها و ابعاد بردار رابطه قابل مشاهده‌ای ندارند و برای انواع دسته‌بندی‌ها بازه ۱۰۰ تا ۲۰۰ مناسب است.

**جدول ۵- نتایج ارزیابی مدل‌ها برای دسته‌بندی ۲ و ۵تایی روی داده‌های مختلف با در نظر گرفتن تاثیر ابعاد بردارهای کلمات**

| ابعاد بردارها | پنج‌تایی | | | دوتایی | | |
|---|---|---|---|---|---|---|
| | نظرات | توئیتر | اخبار | توئیتر | اخبار | نظرات |
| ۱۰ | ۰.۴۱ | ۰.۴۰ | ۰.۳۳ | ۰.۶۱ | ۰.۶۶ | ۰.۶۱ |
| ۵۰ | ۰.۴۳ | ۰.۴۵ | ۰.۴۳ | ۰.۷۵ | ۰.۷۸ | ۰.۷۲ |
| ۱۰۰ | ۰.۴۳ | ۰.۴۳ | ۰.۴۶ | ۰.۷۵ | ۰.۷۹ | ۰.۷۳ |
| ۲۰۰ | ۰.۴۲ | ۰.۴۵ | ۰.۴۴ | ۰.۷۵ | ۰.۷۸ | ۰.۷۳ |
| ۳۰۰ | ۰.۴۲ | ۰.۴۵ | ۰.۴۳ | ۰.۷۴ | ۰.۷۸ | ۰.۷۲ |

پیدا کردن بازه مناسب برای داده‌های آموزش با اندازه بزرگتر از داده حاضر، نیاز به بررسی بازه‌های متفاوت دارد.

## 4-7- تاثیر تعداد تکرار در آموزش بردار کلمات بر نتایج

برای بررسی عدد مناسب تکرار در مرحله آموزش بردارهای کلمات ما ابعاد بردار کلمات را ۱۰۰ و اندازه صافی را سه در نظر می‌گیریم. اعداد تکرار ما ۲۵، ۵۰، ۱۰۰ و ۲۰۰ است.

از نتایج نمایش داده شده در جدول ۶ مشاهده می‌شود که تعداد تکرار بیش‌تر از ۲۵ ارتباط مشخصی با انواع داده با طول‌های مختلف ندارد. همین‌طور نتایج دسته‌بندی‌های مختلف با تعداد دسته مختلف ارتباط معناداری با تعداد تکرار در آموزش بردارها ندارد. تکرار بیش‌تر از ۱۰۰ نتایج را تغییر چندانی نمی‌دهد. بدیهی است که با افزایش تکرار، زمان طی شده برای آموزش بردارها افزایش می‌یابد. تکرار ۱۰۰ نتایج بهتری در مجموع دارد اما با توجه به فاصله بسیار نزدیکی که با تعداد تکرار ۵۰ دارد و با توجه به مدت زمان طی شده برای آموزش به حجم مجموعه ما چندان به‌صرفه به‌نظر نمی‌رسد. اصولا برای زبان‌ها با منابع محدود تعداد تکرار بین ۵۰ تا ۱۰۰ پیشنهاد می‌شود.

**جدول ۶- نتایج ارزیابی مدل‌ها برای دسته‌بندی ۲ و ۵تایی روی داده‌های مختلف با در نظر گرفتن تاثیر تعداد تکرار در آموزش بردار کلمات**

| تعداد تکرار آموزش | پنج‌تایی | | | دوتایی | | |
|---|---|---|---|---|---|---|
| | نظرات | توئیتر | اخبار | توئیتر | اخبار | نظرات |
| ۲۵ | ۰.۳۹ | ۰.۴۰ | ۰.۴۲ | ۰.۷۱ | ۰.۷۲ | ۰.۶۹ |
| ۵۰ | ۰.۴۳ | ۰.۴۳ | ۰.۴۳ | ۰.۷۵ | ۰.۷۸ | ۰.۷۲ |
| ۱۰۰ | ۰.۴۳ | ۰.۴۳ | ۰.۴۶ | ۰.۷۵ | ۰.۷۹ | ۰.۷۳ |
| ۲۰۰ | ۰.۴۴ | ۰.۴۵ | ۰.۴۲ | ۰.۷۵ | ۰.۷۸ | ۰.۷۲ |

## ۵- نتیجه‌گیری و پیشنهادها

در کار حاضر ما شبکه عصبی پیچشی با پارامترهای متفاوت را با استفاده از بردارهای کلمات بر روی رسانه‌های اجتماعی جهت تحلیل احساس متن فارسی به‌کار بردیم. آموزش بردارهای عمومی در شبکه عصبی با یک لایه کانولوشن کارآیی بهتری نسبت به روش‌های سنتی یادگیری ماشین و شبکه‌های عصبی بازگشتی به‌خصوص بر روی داده‌ها با طول کوتاه نشان داد. نتایج ما نشان داد که بردارهای کلمات استخراج شده از داده‌های عمومی بدون توجه به نوع کاربری می‌توانند به بهبود نتایج در پردازش زبان طبیعی کمک کنند. برای تحقیقات آتی، در نظر گرفتن بردارهای کارویژه پیشنهاد می‌شود. همچنین در کار حاضر، ما تنها به دسته‌بندی در سطح جملات پرداختیم. برای کارهای آینده می‌توان به سطوح بالاتر از جمله، مثل کل سند و تحلیل احساس بر اساس موجودیت‌های مختلف، که منجر به تحلیل دقیق‌تر برای جملات و اسناد دارای نظرات با جهت‌گیری متفاوت می‌شود، پرداخت.

## مراجع

[1] Greenwood, S., A. Perrin, and M. Duggan. "Social media update 2016: Facebook usage and engagement is on the rise, while adoption of other platforms holds steady." *Pew Research Center* (2016).

[2] Mander, Jason. "Daily time spent on social networks rises to 1.72 hours." *London: Global Web Index* (2015).

[3] Liu, Bing. "Sentiment analysis and opinion mining." *Synthesis lectures on human language technologies* 5, no. 1 (2012): 1-167.

[4] Pang, Bo, Lillian Lee, and Shivakumar Vaithyanathan. "Thumbs up?: sentiment classification using machine learning techniques." In *Proceedings of the ACL-02 conference on Empirical methods in natural language processing-Volume 10*, pp. 79-86. Association for Computational Linguistics, 2002.

[5] Tripathy, Abinash, Ankit Agrawal, and Santanu Kumar

Kuksa. "Natural language processing (almost) from scratch." *Journal of Machine Learning Research* 12, no. Aug (2011): 2493-2537.

[20] Wang, Keze, Xiaolong Wang, Liang Lin, Meng Wang, and Wangmeng Zuo. "3D human activity recognition with reconfigurable convolutional neural networks." In *Proceedings of the 22nd ACM international conference on Multimedia*, pp. 97-106. ACM, 2014.

[21] Mikolov, Tomas, Ilya Sutskever, Kai Chen, Greg S. Corrado, and Jeff Dean. "Distributed representations of words and phrases and their compositionality." In *Advances in neural information processing systems*, pp. 3111-3119. 2013.

[22] Bengio, Yoshua, Réjean Ducharme, Pascal Vincent, and Christian Jauvin. "A neural probabilistic language model." *Journal of machine learning research* 3, no. Feb (2003): 1137-1155.

[23] Bojanowski, Piotr, Edouard Grave, Armand Joulin, and Tomas Mikolov. "Enriching word vectors with subword information." *arXiv preprint arXiv:1607.04606* (2016).

[24] Bagheri, Ayoub, and Mohamad Saraee. "Persian Sentiment Analyzer: A Framework based on a Novel Feature Selection Method." *International Journal of Artificial Intelligence™* 12, no. 2 (2014): 115-129.

[25] Pedregosa, Fabian, Gaël Varoquaux, Alexandre Gramfort, Vincent Michel, Bertrand Thirion, Olivier Grisel, Mathieu Blondel et al. "Scikit-learn: Machine learning in Python." *Journal of Machine Learning Research* 12, no. Oct (2011): 2825-2830.

[26] Bergstra, James, Olivier Breuleux, Frédéric Bastien, Pascal Lamblin, Razvan Pascanu, Guillaume Desjardins, Joseph Turian, David Warde-Farley, and Yoshua Bengio. "Theano: A CPU and GPU math compiler in Python." In *Proc. 9th Python in Science Conf*, pp. 1-7. 2010.

[27] Hajmohammadi, Mohammad Sadegh, and Roliana Ibrahim. "A svm-based method for sentiment analysis in persian language." In International Conference on Graphic and Image Processing (ICGIP 2012), vol. 8768, p. 876838. International Society for Optics and Photonics, 2013.

[28] Basiri, Mohammad Ehsan, Ahmad Reza Naghsh-Nilchi, and Nasser Ghassem-Aghaee. "A framework for sentiment analysis in persian." Open Transactions on Information Processing 1, no. 3, pp 1-14, 2014.

[29] Roshanfekr, Behnam, Shahram Khadivi, and Mohammad Rahmati. "Sentiment analysis using deep learning on Persian texts." In 2017 Iranian Conference on Electrical Engineering (ICEE), pp. 1503-1508. IEEE, 2017.

[۳۰] اکبریان، حسین، مصطفی صالحی، و هادی ویسی. ۱۳۹۵. تعیین جهت گیری نظرات در رسانه‌های اجتماعی فارسی زبان. ارائه شده در بیست و چهارمین کنفرانس مهندسی برق ایران، شیراز: دانشگاه شیراز.

Rath. "Classification of sentiment reviews using n-gram machine learning approach." *Expert Systems with Applications* 57 (2016): 117-126.

[6] Mullen, Tony, and Nigel Collier. "Sentiment Analysis using Support Vector Machines with Diverse Information Sources." In *EMNLP*, vol. 4, pp. 412-418. 2004.

[7] Agarwal, Basant, and Namita Mittal. "Machine learning approach for sentiment analysis." In *Prominent Feature Extraction for Sentiment Analysis*, pp. 21-45. Springer International Publishing, 2016.

[8] Poria, Soujanya, Haiyun Peng, Amir Hussain, Newton Howard, and Erik Cambria. "Ensemble application of convolutional neural networks and multiple kernel learning for multimodal sentiment analysis." *Neurocomputing* (2017).

[9] Dong, Li, Furu Wei, Chuanqi Tan, Duyu Tang, Ming Zhou, and Ke Xu. "Adaptive Recursive Neural Network for Target-dependent Twitter Sentiment Classification." In *ACL (2)*, pp. 49-54. 2014.

[10] Kim, Yoon. "Convolutional neural networks for sentence classification." *arXiv preprint arXiv:1408.5882* (2014).

[11] Zhang, Ye, and Byron Wallace. "A sensitivity analysis of (and practitioners' guide to) convolutional neural networks for sentence classification." *arXiv preprint arXiv:1510.03820* (2015).

[12] Jaynes, Edwin T. "Information theory and statistical mechanics." *Physical review* 106, no. 4 (1957): 620.

[13] Neethu, M. S., and R. Rajasree. "Sentiment analysis in twitter using machine learning techniques." In *Computing, Communications and Networking Technologies (ICCCNT), 2013 Fourth International Conference on*, pp. 1-5. IEEE, 2013.

[14] Cortes, Corinna, and Vladimir Vapnik. "Support-vector networks." *Machine learning* 20, no. 3 (1995): 273-297.

[15] Dos Santos, Cícero Nogueira, and Maira Gatti. "Deep Convolutional Neural Networks for Sentiment Analysis of Short Texts." In *COLING*, pp. 69-78. 2014.

[16] Zhang, Yu, Mengdong Chen, Lianzhong Liu, and Yadong Wang. "An effective convolutional neural network model for Chinese sentiment analysis." In *AIP Conference Proceedings*, vol. 1836, no. 1, p. 020085. AIP Publishing, 2017.

[17] Cieliebak, Mark, Jan Deriu, Dominic Egger, and Fatih Uzdilli. "A Twitter Corpus and Benchmark Resources for German Sentiment Analysis." *SocialNLP 2017* (2017): 45.

[18] Socher, Richard, Alex Perelygin, Jean Y. Wu, Jason Chuang, Christopher D. Manning, Andrew Y. Ng, and Christopher Potts. "Recursive deep models for semantic compositionality over a sentiment treebank." In *Proceedings of the conference on empirical methods in natural language processing (EMNLP)*, vol. 1631, p. 1642. 2013.

[19] Collobert, Ronan, Jason Weston, Léon Bottou, Michael Karlen, Koray Kavukcuoglu, and Pavel

[31] حسینی، پدرام، علی احمدیان رامکی، حسن ملکی، منصوره انواری، و سید ابولقاسم میرروشندل. ۱۳۹۳. پیکره فارسی تحلیل احساس سنتی پرس. ارائه شده در سومین همایش ملی زبان‌شناسی رایانشی، تهران: دانشگاه صنعتی شریف.

[32] زارع چاهوکی، محمدعلی، و سید حمیدرضا محمدی. ۱۳۹۵ بهینه‌سازی هسته‌های چندگانه ور ماشین‌بردارپشتیبان جفتی برای کاهش شکاف معنایی تشخیص صفحات فریب‌آمیز. مجله مهندسی برق دانشگاه تبریز ۴۶(۴): ۱۳۵-۱۴۵.

[33] رفان، محمدحسین، مهرنوش کمرزرین، و عادل دمشقی. ۱۳۹۵. بهبود دقت و پایداری RTDGPS با استفاده از مدل ترکیبی RNN و PSO. مجله مهندسی برق دانشگاه تبریز ۴۶(۱): ۱۸۵-۱۹۶.

**زیرنویس‌ها**

[1] Subjective Opinions
[2] Objective Opinions
[3] Supervised Learning
[4] Generic
[5] Convolutional Neural Networks
[6] Feedforward
[7] Word Embedding
[8] Naïve Bayes classifier
[9] Maximum Entropy Classifier (MEC)
[10] Exponential Model
[11] Principle of Maximum Entropy
[12] Support Vector Machines (SVMs)
[13] Recurrent Neural Networks
[14] Semantic Role Labeling
[15] Chunking
[16] Representation Learning
[17] Unsupervised Learning
[18] Grid
[19] Filters
[20] Feature Maps
[21] Max-pooling
[22] Bias
[23] Channel
[24] Sentence Vector
[25] Softmax
[26] Global Feature Vector
[27] Distributed Representation of Words
[28] Validation Loss
[29] Dropout
[30] Adadelta
[31] Area Under Curve